\documentclass{mn2e}

\usepackage{epsf}

\title[The massive cluster lens MACS\,J1206.2$-$0847]{A spectacular
  giant arc in the massive cluster lens
  MACS\,J1206.2$-$0847\footnote{Based on observations collected with
    the VLT-UT3 Melipal Telescope (ESO)}}

\author[H.\ Ebeling et al.]{H.\ Ebeling$^{1}$,
  C.J.\ Ma$^{1}$, J.-P.\ Kneib$^{2}$,
  E.\ Jullo$^{2}$, N.J.D.\ Courtney$^{3}$,
  E.\ Barrett$^{1}$, A.C.\ Edge$^{3}$, J.-F.\ Le Borgne$^{4}$
  \\
$^{1}$ Institute for Astronomy, University of Hawaii, 2680
Woodlawn Drive, Honolulu, HI 96822, USA\\
$^{2}$ Laboratoire d'Astrophysique de Marseille, OAMP, CNRS-Universit\'e Aix-Marseille, 38 rue Fr\'ed\'eric Joliot-Curie, 13388 Marseille Cedex 13, France\\
$^{3}$ Department of Physics, University of Durham, South Road, Durham, DH1 3LE, UK\\
$^{4}$ Laboratoire d'Astrophysique de Toulouse-Tarbes, Universit\'{e} de Toulouse, CNRS, 14 Avenue Edouard Belin, F-31400 Toulouse, France}

\begin{document}

\date{Accepted 2008 TBD. Received 2008 TBD; in original form 2008 TBD}

\pagerange{\pageref{firstpage}--\pageref{lastpage}} \pubyear{2008}

\maketitle

\label{firstpage}

\begin{abstract}
  We discuss the X-ray and optical properties of the massive galaxy cluster
  MACS\,J1206.2$-$0847 ($z=0.4385$), discovered in the Massive Cluster Survey
  (MACS). Our Chandra observation of the system yields a total X-ray luminosity
  of $2.4\times 10^{45}$ erg s$^{-1}$ (0.1--2.4 keV) and a global gas
  temperature of $11.6\pm 0.7$ keV, very high values typical of MACS
  clusters. In both optical and X-ray images MACS\,J1206.2$-$0847 appears close
  to relaxed in projection, with a pronounced X-ray peak at the location of the
  brightest cluster galaxy (BCG); we interpret this feature as the remnant of a
  cold core. A spectacular giant gravitational arc, 15\arcsec in length, bright
  (V$\sim21$) and unusually red ($R-K=4.3$), is seen 20 arcsec west of the BCG;
  we measure a redshift of $z=1.036$ for the lensed galaxy. From our Hubble
  Space Telescope image of the cluster we identify the giant arc and its counter
  image as a seven-fold imaged system.  An excess of X-ray emission in the
  direction of the arc coincides with a mild galaxy overdensity and could be the
  remnant of a minor merger with a group of galaxies. We derive estimates of the
  total cluster mass as well as of the mass of the cluster core using X-ray,
  dynamical, and gravitational-lensing techniques. For the mass enclosed by the
  giant arc ($r<119$ kpc) our strong-lensing analysis based on Hubble Space
  Telescope imaging yields a very high value of $1.1\times10^{14}$ M$_{\odot}$,
  inconsistent with the much lower X-ray estimate of $0.5\times10^{14}$
  M$_{\odot}$. Similarly, the virial estimate of $4\times10^{15}$ M$_{\odot}$
  for the total cluster mass, derived from multi-object spectroscopy with CFHT
  and the VLT of 38 cluster members, is significantly higher than the
  corresponding X-ray estimate of $1.7\times10^{15}$ M$_{\odot}$. We take the
  discrepancy between X-ray and other mass estimates to be indicative of
  pronounced substructure along the line of sight during an ongoing merger
  event, an interpretation that is supported by the system's very high velocity
  dispersion of 1580 km s$^{-1}$.
\end{abstract}


\section{Introduction}
The concentration of both dark and baryonic matter in the cores of
clusters of galaxies has many profound implications for our
understanding of cluster growth and cosmology. Firstly, the structure
and evolution of the gravitational potential of a cluster of galaxies
depends on the nature of dark matter and thus allows direct comparison
with predictions from numerical simulations (e.g.\ Navarro, Frenk \&
White 1997).  Secondly, the surface density of mass integrated through
the core of a cluster is often sufficiently high to strongly lens
background galaxies into gravitational arcs (Soucail et al.\ 1987;
Mellier et al.\ 1991; Kneib et al.\ 1996; Smith et
al.\ 2001). Detailed analysis of the location and shape of such arcs,
as well as of lens-generated multiple images, can be used to model the
projected mass in the cluster, and in-depth follow-up of the brightest
lensed features often yields valuable insights into the properties of
distant, faint galaxies (e.g.\ Kneib et al.\ 2004; Smail et
al.\ 2007).  Thirdly, the high density and temperature of the gas in
the core of clusters leads to intense X-ray emission that current
instrumentation can detect out to redshifts well above unity. The
selection of massive clusters through X-ray emission has proved very
successful at providing cosmological constraints (Henry 2000; Borgani
et al.\ 2001; Allen et al.\ 2003; Pierpaoli et al.\ 2003; Allen et
al.\ 2008; Mantz et al.\ 2008) and follow-up observations of X-ray
luminous clusters have revealed many spectacular cases of
gravitational lensing (Gioia \& Luppino 1994; Smith et al.\ 2001;
Dahle et al.\ 2002; Covone et al.\ 2006).

In this paper we present a comprehensive multiwavelength study of a
complex gravitational arc and its host cluster MACS\,J1206.2$-$0847,
an X-ray selected system at intermediate redshift found by the Massive
Cluster Survey, MACS (Ebeling, Edge \& Henry 2001, 2007). We describe
our optical, NIR and X-ray observations in Sections 2 and 3,
investigate the properties of the giant arc, the cluster lens, and the
brightest cluster galaxy in Section 4 to 7, and derive mass estimates
for the cluster core and the entire system in Section 7. We present a
discussion of our results as well as conclusions in Section 8.

Throughout we use a $\Lambda$CDM cosmology ($\Omega_M=0.3$,
$\Omega_{\lambda}=0.7$) and adopt $H_0=$ 70 km s$^{-1}$ Mpc$^{-1}$.

\section{Observations}

The galaxy cluster MACS\,J1206.2$-$0847 was originally discovered in a
short two-minute R-band image taken on June 15, 1999 with the
University of Hawaii's 2.2m telescope (UH2.2m) on Mauna Kea. The
observation, performed as part of the MACS project, was triggered by
the presence of the X-ray source 1RXS\,J120613.0$-$084743 in the ROSAT
Bright Source Catalogue which had no obvious counterpart in the
standard astronomical databases and could also not trivially
be identified by inspection of the respective Digitized Sky Survey
image.

Following the initial, tentative identification of the RASS X-ray
source as a potentially massive galaxy cluster, we conducted a range
of follow-up observations to firmly establish the cluster nature of
this source and to characterize its physical properties.

\subsection{Optical}
\label{optspec}

Spectra of two galaxies in MACS\,J1206.2$-$0847, one of them the BCG,
were taken with the Wide-Field Prism Spectrograph on the UH2.2m on
July 4, 1999, using a 420 l/mm grism, a Tektronix 2048$^2$ CCD
yielding 0.355 arcsec/pixel, and a 1.6 arcsec slit. The two redshifts
were found to be concordant, establishing an approximate cluster
redshift of $z\approx 0.434$.

Moderately deep, multi-passband imaging observations ($3\times 240$s,
dithered by 10 arcsec, in each of the V, R, and I filters) of the
cluster were obtained with the UH2.2m on January 29, 2001, using again
the Tektronix 2048$^2$ CCD which provides a scale of 0.22 arcsec per
pixel and a $7.3\times7.3$ arcminute$^2$ field of view. The seeing was
variable throughout the night; we measure seeing values of 0.85, 1.05,
0.90 arcsec in the V, R, and I passbands, respectively, from the
final, co-added images.

Spectroscopic observations of presumed cluster galaxies as well as of
the giant arc in MACS\,J1206.2$-$0847 were performed with the FORS1
spectrograph in multi-object spectroscopy mode at the UT3 Melipal
telescope of the VLT on April 11, 2002. The G300V grism, an order
sorting filter (GG375), and a 1 arcsec slit were used, yielding a
wavelength coverage from $\sim$4000\AA\ to $\sim$8600\AA\ at a
resolution of $R=500$. The total exposure time was 38 minutes. A
single mask was designed, covering the $\sim$7' field of FORS1 with 19
slitlets of fixed length (22"). Credible redshifts could be measured
for 14 objects. During the exposure the seeing was 0.6 arcsec. Spectra
of the spectrophotometric standard star EG274 were obtained for
calibration.

Additional multi-object spectroscopy of colour-selected galaxies in
the field of MACS\,J1206.2$-$0847 was performed on May 8, 2003 using
the MOS spectrograph on the Canada-France-Hawaii Telescope (CFHT) on
Mauna Kea. We used the B300 grism and the EEV1 CCD, which provides a
resolution of 3.3\AA/pixel, and observed through a broadband filter
(\#4611, General Purpose set) to produce truncated spectra, covering
about 1800\AA\ centred on 6150\AA, such that spectra could be stacked
in three tiers along the dispersion direction. This choice of filter
and grism ensured that, for all cluster members, redshifts could be
obtained from the Ca H+K lines which fall at 5660\AA\ and 5710\AA\ at
the approximate cluster redshift of $z=0.44$. Weather conditions were
poor though, and only 48 of the 67 objects observed (total integration
time: one hour) yielded reliable redshifts.

Finally, MACS\,J1206.2$-$0847 was observed on December 6, 2005 with
the Advanced Camera for Surveys (ACS) aboard the Hubble Space
Telescope as part of program SNAP-10491 (PI Ebeling), for a total of
1200 seconds in the F606W filter, resulting in a high-resolution image
of the cluster core, including the giant arc.

\subsection{Near-infrared}

Near-infrared observations of the core of MACS\,J1206.2$-$0847 were
performed in the J and K bands using the United Kingdom Infra-Red
telescope (UKIRT) on April, 5 2001 using the UFTI imager during a
period of good seeing. The observations consisted of two iterations of
a nine-point dither pattern, each of 60 second exposures for a total
integration time of 1080 seconds. The seeing measured from these
observations was 0.47 and 0.59 arc-seconds in the J and K bands,
respectively.

\subsection{X-ray}

MACS\,J1206.2$-$0847 was observed on December 18, 2002 with the ACIS-I
detector aboard the Chandra X-ray Observatory for a nominal duration
of 23.5 ks, as part of a Chandra Large Programme awarded to the MACS
team. The target was placed about two arcmin off the standard aimpoint
to avoid flux being lost in the chip gaps of the ACIS-I detector, while
still maintaining good (sub-arcsec) angular resolution across the
cluster core. VFAINT mode was used in order to maximize the efficiency
of particle event rejection in the post-observation processing.

\section{Data reduction}

\subsection{Optical imaging}

Standard data-reduction techniques (bias-subtracting, flat-fielding,
image combination and registration) were applied to the V, R and I
band data using the relevant \textsc{iraf} packages. The data were
photometrically calibrated via the observation of Landolt
standard-star fields.

In order to measure aperture magnitudes, seeing-matched frames were
produced by applying a Gaussian smoothing to the V, I, J and K images
such that the seeing in these frames was degraded to match the 1.1
arc-seconds measured in the R band. Using SExtractor, the
seeing-matched (undegraded) images were then used to measure aperture
(total) magnitudes.

Photometry of the giant arc was performed by manually defining an
aperture mask fitted to the arc profile. The area defined by this
aperture was then masked out of the science frame and a background
image produced by median smoothing over the absent arc. This
background image was subtracted from the science frame and VRIJK
photometry obtained by applying the aperture to the resulting
sky-subtracted images.

\subsection{Optical spectroscopy}

For the reduction of our spectroscopic data we applied the same
standard techniques as for the imaging data, followed by straightening
of the skylines, extraction of spectra, and wavelength calibration
using the relevant \textsc{iraf} packages.

Preliminary redshifts were determined via visual inspection, typically
from the calcium H and K lines. Final refined redshifts were found
with a multi-template cross-correlation method using the \textsc{iraf}
task \textsc{fxcor}.

Redshifts with cross-correlation peak heights exceeding 0.70 are
typically accurate to 0.0002; for peak heights between 0.5 and 0.7 the
error is typically 0.0005. All redshifts were converted to the
heliocentric reference frame. Table~\ref{ztbl} lists the coordinates
of all galaxies observed, as well as the measured redshifts with their
uncertainties.

\subsection{Near-infrared imaging}

The J and K imaging data were reduced using the ORAC-DR data-reduction
pipeline and were calibrated through observation of observatory
standard stars.

\begin{figure}
\epsfxsize=0.5\textwidth
\epsffile{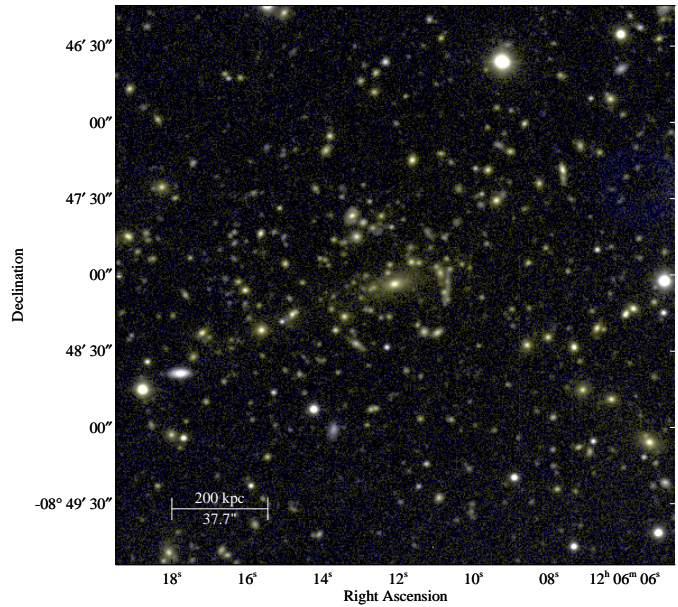}
\caption{\label{imvri} Colour image (IVR mapped to RGB) of
  MACSJ1206.2$-$0847 from $3\times 240$s observations (per filter)
  obtained with the UH 2.2m telescope (see text for observational
  details).}
\end{figure}

\begin{figure}
\epsfxsize=0.5\textwidth
\epsffile{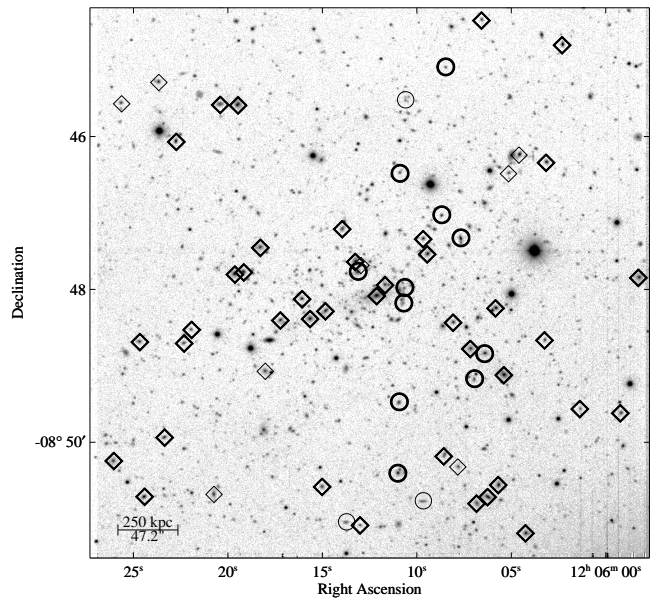}
\caption{\label{imgalpos}Locations of all galaxies for which
  spectroscopic redshifts were measured with the VLT (circles) and
  CFHT (diamonds), overlaid on the UH2.2m R-band image. Bold symbols
  mark galaxies found to be cluster members. See Table~\ref{ztbl} for
  coordinates and redshifts.}
\end{figure}

\subsection{X-ray}

We use {\sc Ciao} (version 3.3), the standard suite of software tools
developed for the analysis of Chandra data at the Chandra Science
Center, as well as the most recent calibration information, to
reprocess the raw ACIS-I data. Our inspection of the lightcurve of the
event count rate in the source-free regions of the ACIS-I detector
finds no significant flaring, leading to an effective (dead-time
corrected) total exposure time of 23.2 ks.

To investigate potential spatial variations in the cluster gas
temperature we define various source and background regions.  For each
of these regions we generate auxiliary response files (ARF) and
response matrix files (RMF) which weigh the position-dependent
instrument characteristics by the observed count distribution in the
respective area. Following Markevitch \& Vikhlinin (2001), only the
0.5--2.0 keV band data are used to create these maps of spatial
weights, since the effects of vignetting are small in this energy
range. We also apply a correction of a factor of 0.93 to the effective
area distribution at energies below 2 keV in all ARFs as suggested by
a comparison of calibration results for the front- and
back-illuminated ACIS chips (Vikhlinin et al.\ 2002). Finally, we use
the {\sc acisabs} package to correct all ARFs for the effects of the
(time-dependent) buildup of a contaminating deposit on the optical
detector window which results in a reduction of the effective area at
low energies.

Background regions are defined by copying the respective source
regions to the same chip-y location on the other three ACIS-I CCDs.
This strategy minimizes the impact of any residual chip-y dependence
of the background on the data analysis, an effect that is unavoidable
if the background is selected as a source-free region on the same CCD
as the cluster.

\section{Cluster galaxy distribution}
\label{galprop}

Figure~\ref{imvri} shows a colour image of the cluster generated from
the V, R, I images obtained with the UH2.2m. MACS\,J1206.2$-$0847 is
found to be an optically very rich system with a single dominant
central galaxy and no obvious subclustering in the apparent, projected
cluster galaxy distribution. A giant arc is clearly visible about 20
arcsec to the West of the BCG. The astrometric solution used is based
on eight stars within the field of view of the V band image that have
accurate celestial coordinates in the Hubble Guide Star Catalogue
(GSC2).

\begin{figure}
\epsfxsize=0.5\textwidth
\epsffile{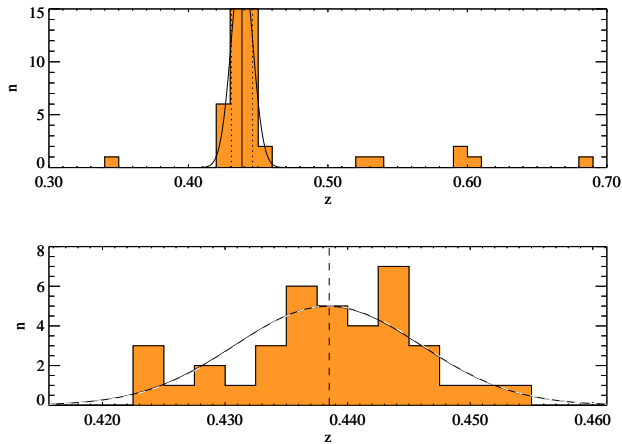}
\caption{\label{zhist}Histogram of galaxy redshifts in the field of
  MACS\,J1206.2$-$0847 as observed with multi-object spectrographs on
  CFHT and the VLT (cf.\ Table~\ref{ztbl}). The overlaid Gaussian
  curve is characterized by the best-fit values for the systemic
  redshift and cluster velocity dispersion of $z=0.4385$ and
  $\sigma=1575^{+191}_{-190}$ km/s, respectively. }
\end{figure}

Our spectroscopic observations of 85 galaxies in the field of
MACS\,J1206.2$-$0847 (Fig.~\ref{imgalpos}) yield redshifts as listed
in Table~\ref{ztbl}.  Two galaxies were observed with both the VLT and
CFHT; their spectroscopic redshifts agree within the errors. Using
only the most accurate redshifts with correlation peak heights
exceeding 0.7, we apply iterative $3\sigma$ clipping to the redshift
histogram to obtain a systemic cluster redshift of $z=0.4384$ and a
very high velocity dispersion in the cluster rest frame of
$\sigma=1581$ km/s based on 38 redshifts. Entirely consistent values
of $z=0.4385$ and $\sigma=1575^{+191}_{-190}$ km/s are found using the
ROSTAT statistics package (Beers, Flynn \& Gebhardt 1990). The
resulting redshift histogram is shown in Fig.~\ref{zhist}. Despite the
extremely high velocity dispersion of the system we find no obvious
signs of substructure along the line of sight; a one-sided
Kolmogorov-Smirnov test finds the observed redshift distribution to be
only mildly inconsistent with a Gaussian ($2.05\sigma$ significance).

\section{Arc properties}

\begin{figure*}[h]
\vspace*{-0.1cm}
\parbox{0.79\textwidth}{
\epsfxsize=0.2\textwidth
\hspace*{9cm} \epsffile{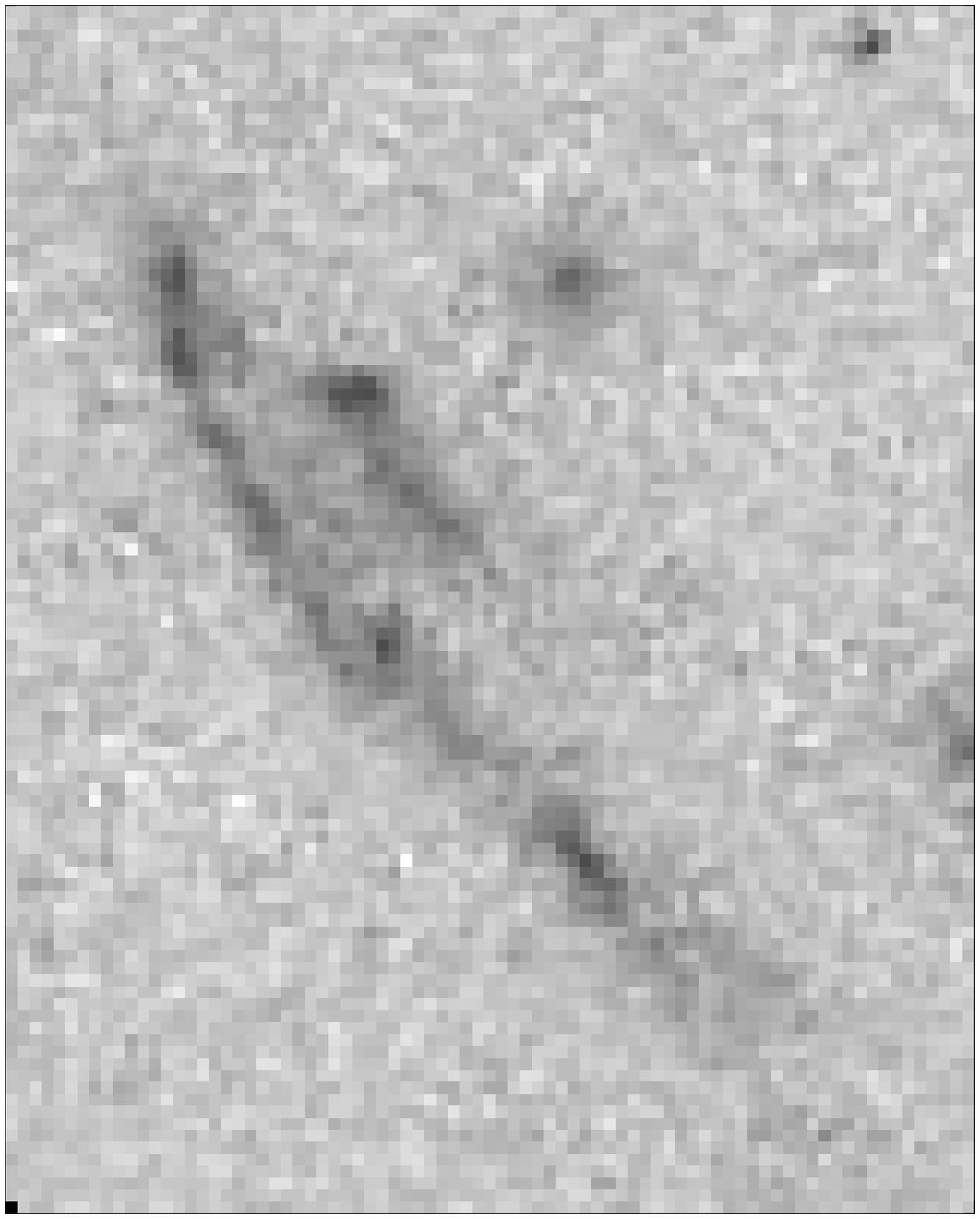}\\
\epsfxsize=0.785\textwidth
\hspace*{5mm}\epsffile{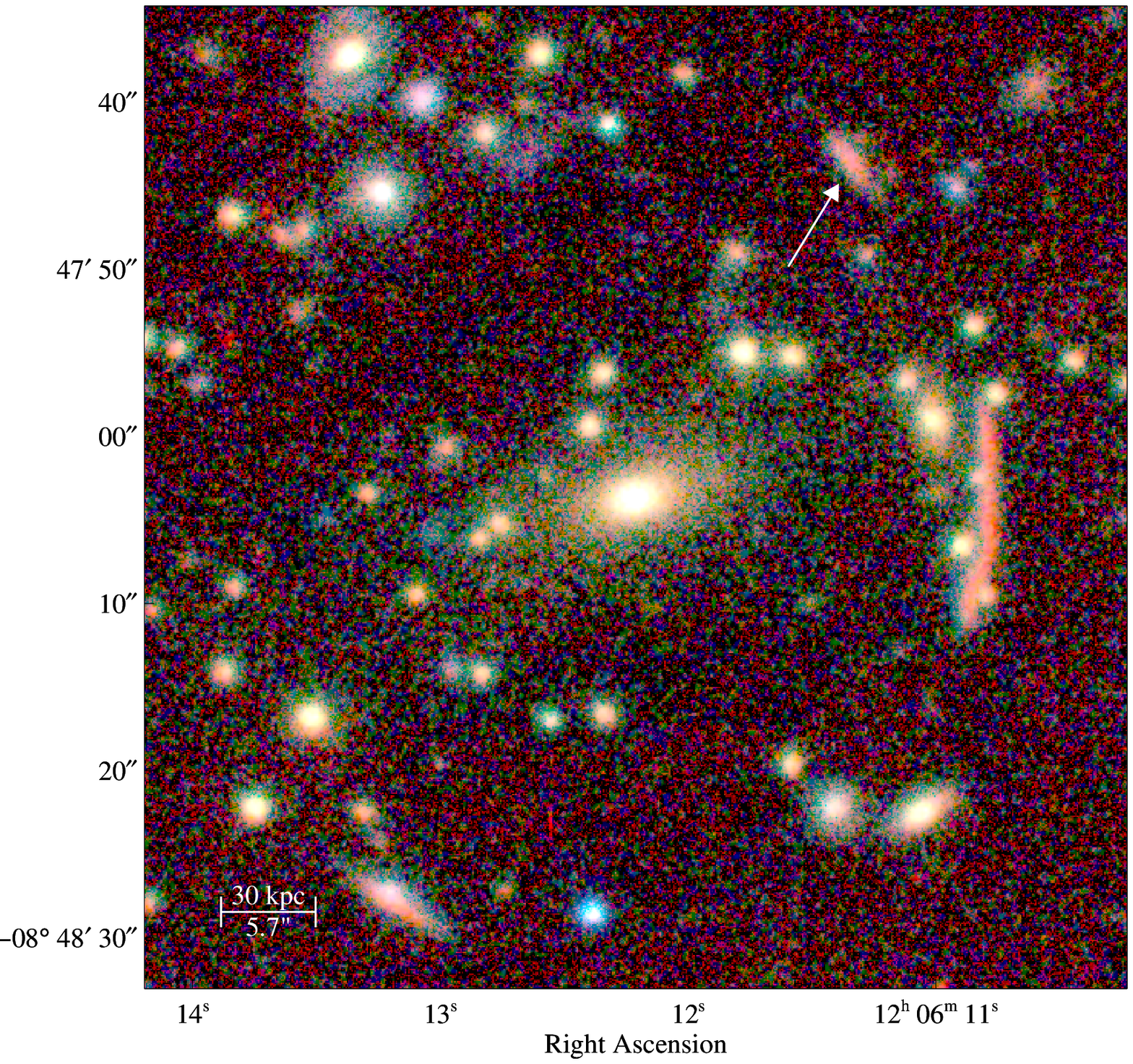}}
\parbox[c]{0.18\textwidth}{
\vspace*{1.8cm}
\epsfxsize=0.2\textwidth
\epsffile{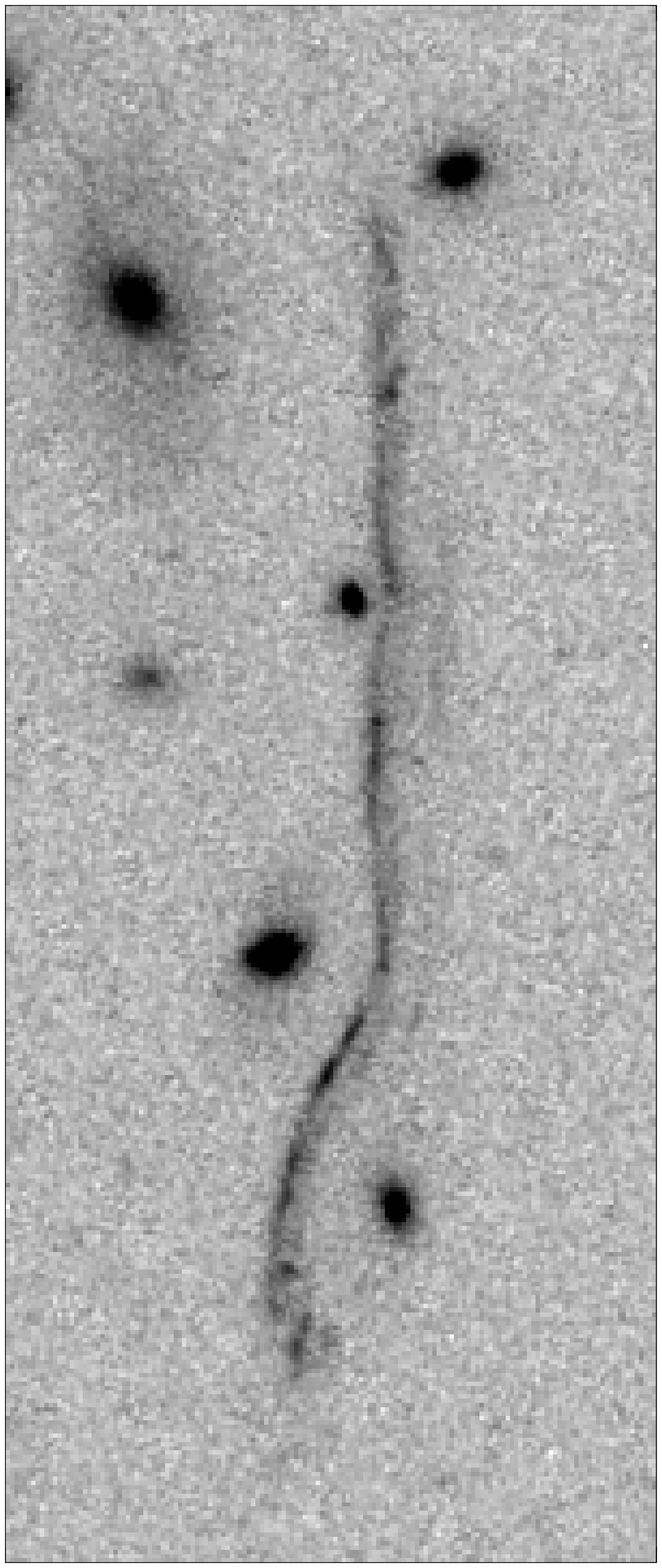}}\mbox{}\\*[-0.1cm]
\caption{\label{imvik}Colour image (KIV mapped to RGB) of MACSJ1206.2$-$0847
  from $3\times 240$s observations (I,V) and $18\times 60$s observations (K)
  obtained with the UH2.2m and UKIRT, respectively (see text for observational
  details). An arrow points to the counter image of two of the four background
  galaxies distorted by the cluster's gravitational field to create the
  prominent giant arc 20 arcsec west of the BCG. Smaller panels on top and to
  the right show enlarged high-resolution views of the arc and its counter image
  as observed with HST/ACS in a 1200s snapshot with the F606W filter. }
\end{figure*}

\begin{figure}
\epsfxsize=0.5\textwidth
\epsffile{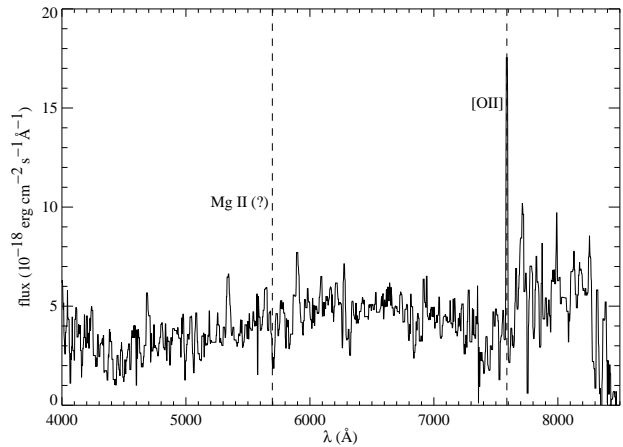}
\caption{\label{arcspec}Spectrum of the giant gravitational arc in
  MACS\,J1206.2$-$0847 as observed with FORS1 on the VLT. We interpret
  the single emission line observed at 7589\AA\ as being [O II], which
  translates into a redshift of 1.036 for the lensed galaxy. Our
  measurement is confirmed by independent observations conducted
  almost simultaneously by Sand et al.\ (2003) using the ESI echelle
  spectrograph on Keck-II which resolves the [O II] doublet thereby
  making the identification unambiguous.}
\end{figure}

The bright ($V=21.0$) giant arc is centered at $\alpha$ (J2000) =
12$^{\rm h}$ 06$^{\rm m}$ 10.75$^{\rm s}$, $\delta$ (J2000) =
$-$08$^{\circ}$ 48' 04.5'', about 20\arcsec west of the BCG.  Its
unusually red colour is apparent in Fig.~\ref{imvik} which shows a
composite V, I, K image of the cluster core. Also prominent, and
marked by the arrow in Fig.~\ref{imvik}, is the counter image, clearly
identifiable by its colour, at $\alpha$ (J2000) = 12$^{\rm h}$
06$^{\rm m}$ 11.27$^{\rm s}$, $\delta$ (J2000) = $-$08$^{\circ}$ 47'
43.0''. Enlarged high-resolution views of arc and counter image
provided by HST are shown in the margins of Fig.~\ref{imvik}. The
photometric properties of arc and counter image are summarized in
Table~\ref{arctbl}; note however that at the resolution of our
ground-based images both the arc and its counter image is blend of
several objects. Figure~\ref{arcspec} shows the spectrum of the giant
arc in MACS\,J1206.2$-$0847 as observed with FORS1 on the VLT.

\begin{table*}
\centering
\caption{Extinction-corrected magnitudes and colours of the
giant arc in MACSJ1206.2$-$0847 and its counter image
(cf.\ Fig.~\ref{imvik}). \label{arctbl}}
\begin{tabular}{@{}lccccccccccc@{}}
\hline
object & I & R & V & J & K & R$-$I & V$-$I & V$-$R & R$-$J & R$-$K & J$-$K\\
\hline
giant arc    & $19.27\pm 0.07$ & $20.27\pm 0.11$ & $20.99\pm 0.08$ & $17.82\pm 0.06$ & $15.95\pm 0.06$ & $1.00\pm 0.12$ & $1.72\pm 0.10$ & $0.72\pm 0.13$ & $2.45\pm 0.12$ & $4.32\pm 0.12$ & $1.87\pm 0.08$ \\
counter image & $20.41\pm 0.06$ & $21.34\pm 0.07$ & $21.94\pm 0.07$ & $18.88\pm 0.06$ & $17.14\pm 0.06$ & $0.93\pm 0.09$ & $1.52\pm 0.09$ & $0.59\pm 0.10$ & $2.46\pm 0.09$ & $4.20\pm 0.09$ & $1.74\pm 0.09$ \\
\hline
\end{tabular}
\end{table*}

At $R-K=4.3$ the giant arc in MACS\,J1206.2$-$0847 is among the
reddest strongly lensed features currently known. Its colour is
comparable to that of the giant arc in Abell 370 ($R-K=4.1$,
Arag\'{o}n-Salamanca \& Ellis 1990) and only slightly bluer than the
red arc in Abell 2390 ($R-K=4.6$, Smail et al.\ 1993).

\begin{figure}
\epsfxsize=0.5\textwidth
\epsffile{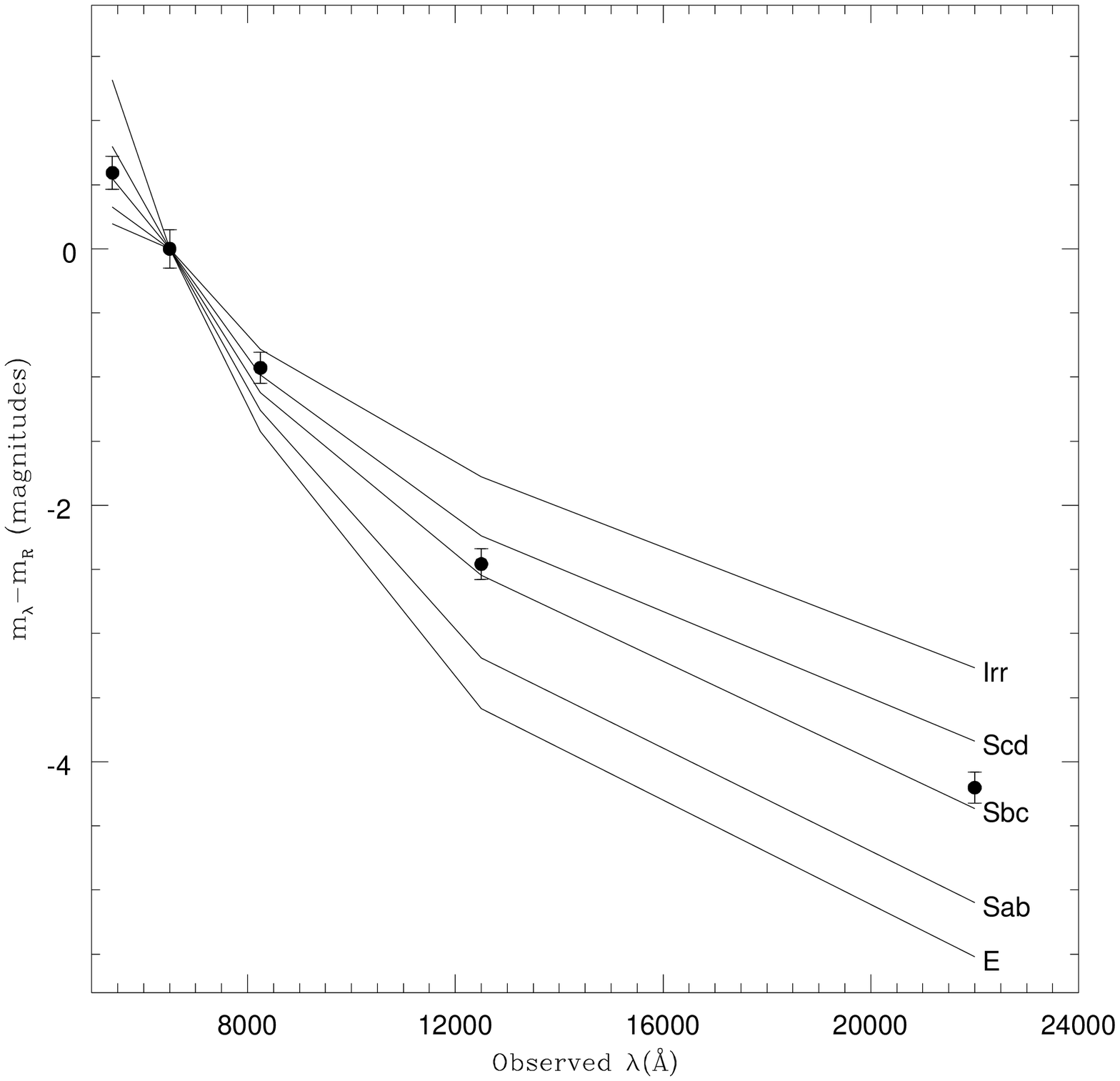}
\caption{\label{arcsed}Broad-band colours of the giant gravitational
  arc in MACS\,J1206.2$-$0847, relative to the R band, compared to
  predicted colours for standard galaxy types at the same
  redshift. The best agreement is found for a spiral of type Sbc to
  Scd.}
\end{figure}

Our multi-band photometry allows the classification of the background
galaxy lensed into the giant arc. Standard template spectral energy
distributions (SEDs) of five galaxy types were redshifted to match
that of the giant arc. The filter response curves of each of the V, R,
I, J and K bands were then convolved with these SEDs and normalized to
the R band to produce predicted colours for each of the five galaxy
types, all relative to the R band. The resulting model
m$_{\lambda}$--m$_{R}$ colours, as well as the equivalent observed
colours of the arc, are shown in Fig.~\ref{arcsed}. We find the colour
distribution of the arc in MACS\,J1206.2$-$0847 to be consistent with
the background galaxy being a normal spiral of class Sbc.

\section{Intra-cluster gas properties}
\label{gas}

\begin{figure}
\epsfxsize=0.5\textwidth
\epsffile{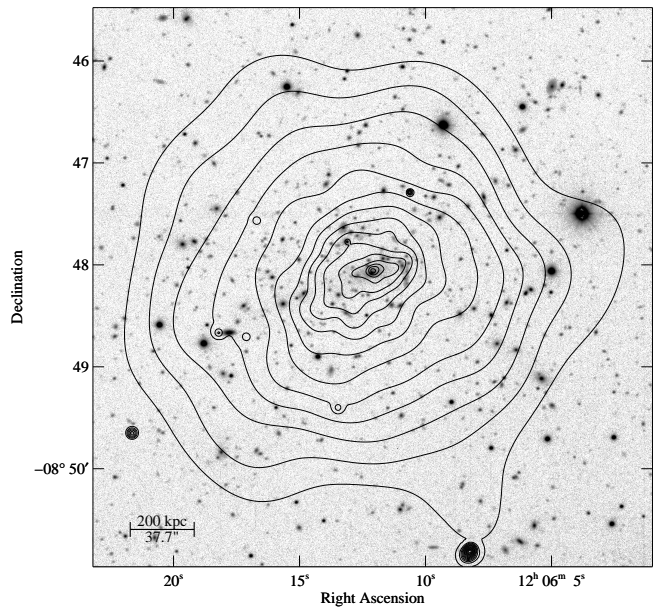}
\caption{\label{xcont}Iso-intensity contours of the adaptively
  smoothed X-ray emission from MACS\,J1206.2$-$0847 in the 0.5--7 keV
  range as observed with Chandra's ACIS-I detector, overlaid on the
  UH2.2m R-band image. The algorithm used, {\textit Asmooth} (Ebeling,
  White \& Rangarajan 2005), adjusts the size of the smoothing kernel
  across the image such that the signal under the kernel has constant
  significance (here 3$\sigma$) with respect to the local background.
  The lowest contour is placed at twice the value of the X-ray
  background in this observation; all contours are logarithmically
  spaced by factors of 1.5.}
\end{figure}

Figure~\ref{xcont} shows contours of the adaptively smoothed X-ray
emission from MACS\,J1206.2$-$0847, as observed with Chandra/ACIS-I in
the 0.5--7 keV band, overlaid on the UH2.2m R-band image\footnote{
For this overlay, as well as for any other comparisons of the spatial
appearance of the cluster in the X-ray and optical wavebands, we have
used three X-ray point sources with obvious optical counterparts to
slightly adjust the astrometric solution of the X-ray image, namely by
$-$0.03 seconds in right ascension and $-$0.4 arcseconds in
declination.  We estimate the resulting, relative astrometry between
the optical and X-ray images to be accurate to better than 0.2
arcseconds.}.

At large distances from the center, the system's X-ray appearance is
close to spherical in projection. The central region at $r\la 250$
kpc, however, shows a pronounced ellipticity as well as non-concentric
X-ray flux contours in the cluster core. The observed elongation, as
well as the displacement of the innermost contours toward the very
compact X-ray core, are both in the direction of a group of galaxies
in the vicinity of the giant arc.

\subsection{Spatial analysis}
\label{xbeta}

Using {\sc Sherpa}, the fitting package provided with {\sc Ciao}, we
fit the observed X-ray surface brightness distribution within 2.5
arcmin of the cluster core with a two-dimensional spatial model,
consisting of an elliptical $\beta$ model
\[
S(r) = S_0 \left[ 1+\left(\frac{r}{r_0}\right)^2\right]^{-3\beta + \frac{1}{2}},
\]
(where $r=r(\phi)$ is the variable radius of an ellipse with
ellipticity $\epsilon$ and orientation angle $\Theta$), an additional
circular Gaussian component to account for the compact core, and a
constant background. 

Point sources with detection significances exceeding 99\% (as measured with the
{\sc celldetect} algorithm) have been excised from the image, and a spectrally
weighted exposure map is used as a two-dimensional response function in the
fit. We use a composite exposure map to account for the differences between
photons of cosmic origin and high-energy particles, the latter being subject
neither to off-axis vignetting nor to variable detection rates due to spatial or
temporal variations in the CCD quantum efficiency (QE). At energies between 0.5
and 2 keV particles account for roughly 2/3 of the observed background; at
higher energies this fraction rises to over 90 per cent. Since, overall, more
than 80\% of the background events registered in the 0.5--2 keV band used here
are caused by particles we ignore the sky contribution altogether and compute a
background exposure map which incorporates the effects of bad pixels and
dithering but not those of a variable CCD QE and vignetting. A second exposure
map is computed using weights based on the spectrum of the target of our
observations. The spectral weights for this `cluster-weighted' exposure map are
created assuming a plasma model with $kT=12$ keV, a metal abundance of 0.3
solar, and the Galactic value of $4.23\times 10^{20}$ cm$^{-2}$ for the
equivalent Hydrogen column density value in the direction of the cluster,
consistent with the results of our spectral fits to the global cluster X-ray
spectrum (see below). The resulting exposure map shows significant vignetting of
more than 20\% across the ACIS-I field of view. The final exposure map used in
the following is then the weighted average of the background and cluster
exposure maps, with the weights being given by the fraction of counts from the
cluster and the background, and with the peak value set to that of the
cluster-weighted map.

Since the image used in the fit is relative coarsely binned ($2\times
2$ arcsec$^2$) and no small-scale spatial components are included in
the model, we do not convolve the model with the telescope
point-spread function (PSF). All parameters of the two-dimensional
spatial model are fit. Because of the low number of counts (zero or
one) in the large majority of image pixels, we use the C statistic
(Cash 1979) during the optimization process.

\begin{figure}
\epsfxsize=0.5\textwidth
\epsffile{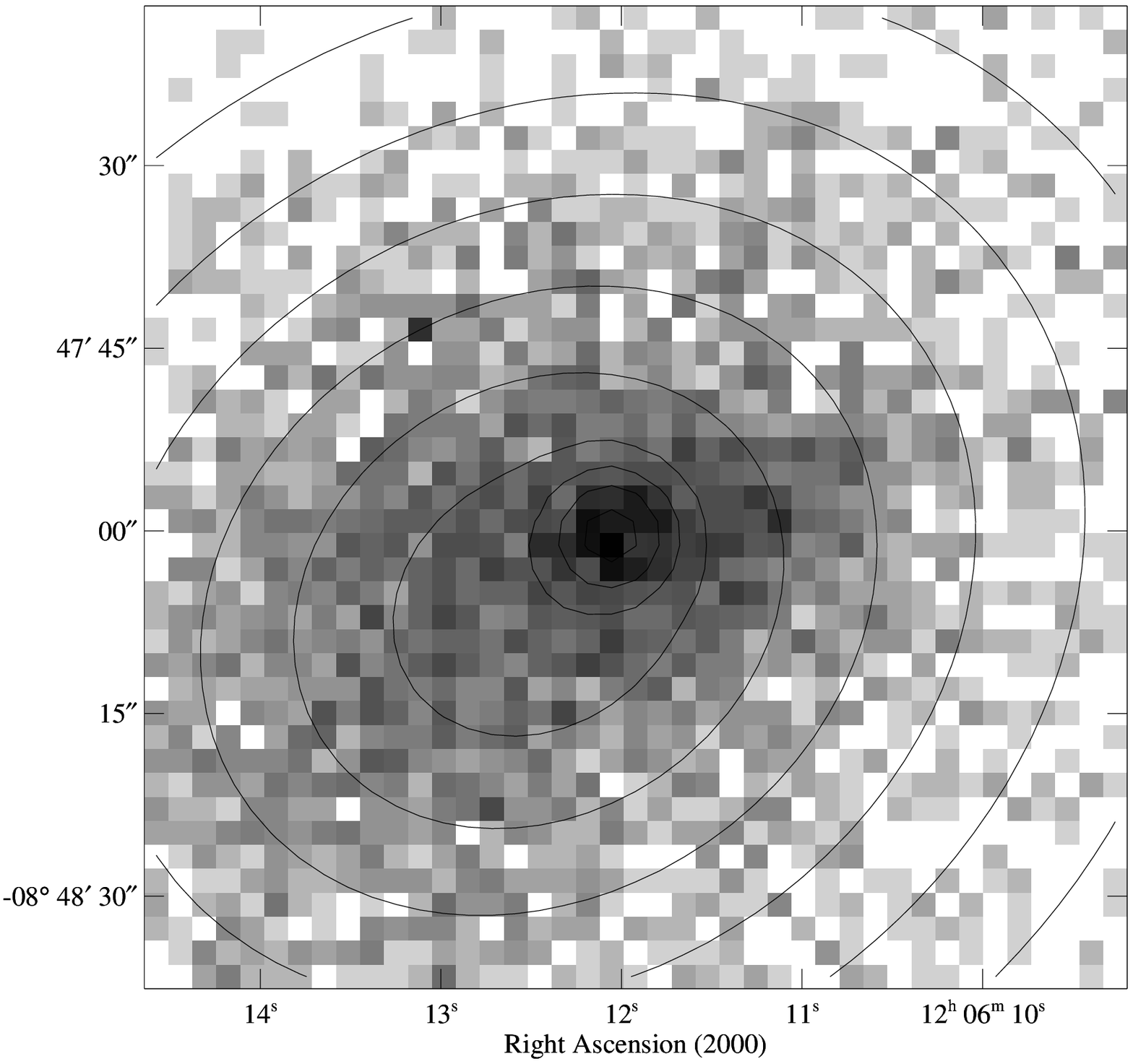}
\epsfxsize=0.5\textwidth
\epsffile{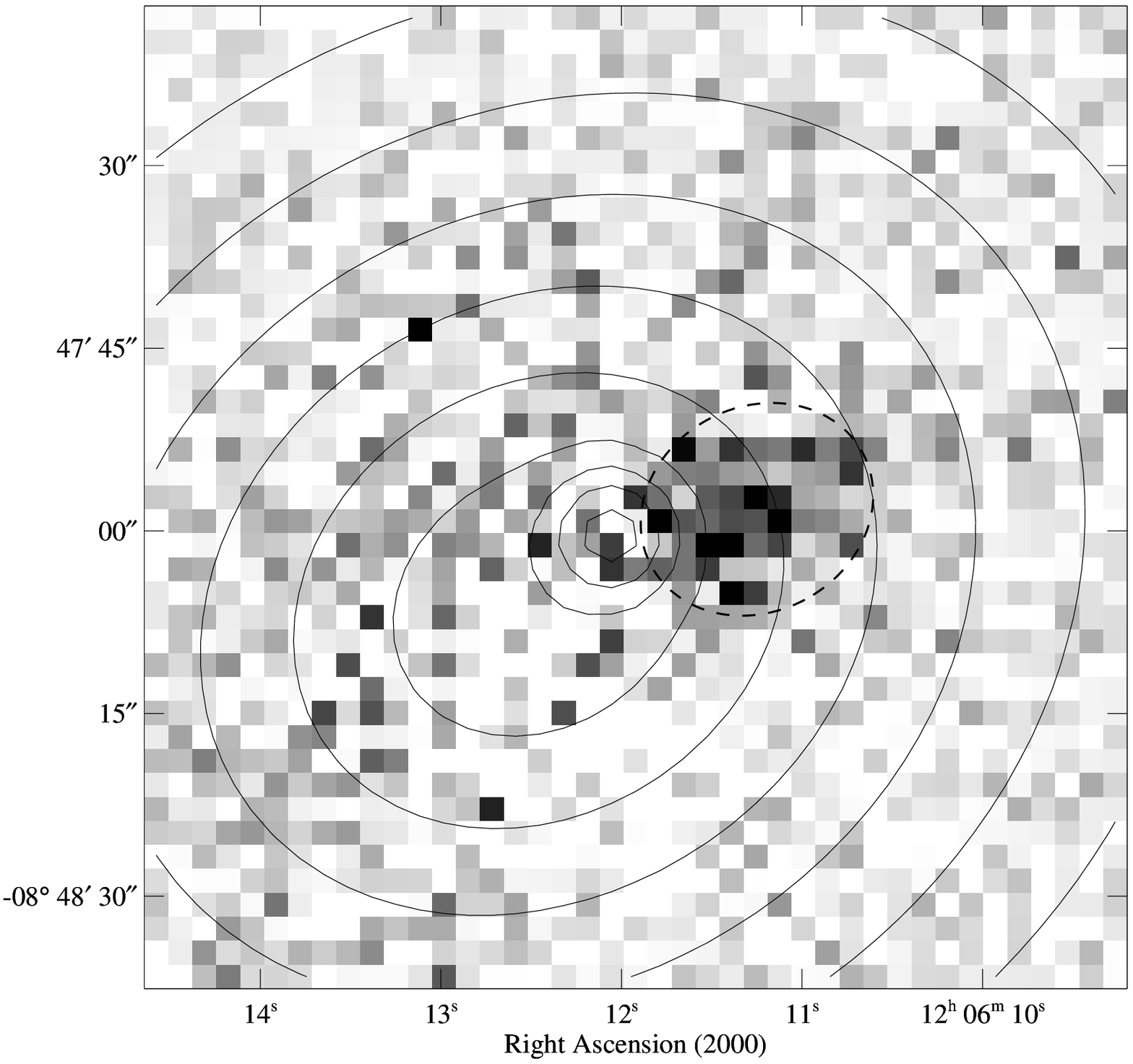}
\caption{\label{2dxfit}Iso-intensity contours (logarithmically spaced)
  of the best-fitting analytic model (elliptical $\beta$ model plus
  circular Gaussian component) of the cluster emission in the 0.5--7
  keV range, overlaid on the observed image (top, logarithmic scaling)
  and the residual image (bottom, linear scaling).  The dashed ellipse
  in the bottom panel highlights the excess emission in the direction
  of the arc and marks the region excluded in the spatial fit of the
  data to the two-dimensional surface-brightness model.}
\end{figure}

Our spatial fit yields best-fit values for the parameters of the $\beta$ model
of $r_0=(23.6\pm 0.8)$ arcsec (corresponding to ($134\pm 5)$ kpc) for the core
radius, $\beta=0.57\pm 0.01$ for the slope parameter, $\epsilon=0.17\pm 0.01$
for the ellipticity, and $\Theta=(56\pm 2)$ degrees (counted North through
West). Our two-dimensional spatial fit also finds coordinates of $\alpha$
(J2000) = 12$^{\rm h}$ 06$^{\rm m}$ 12.10$^{\rm s}$, $\delta$ (J2000) =
$-$08$^{\circ}$ 48' 01.7'' for the centroid of the compact cluster core, offset
by $(8.1\pm 1.7)$ arcsec from the position $\alpha$ (J2000) = 12$^{\rm h}$
06$^{\rm m}$ 12.50$^{\rm s}$, $\delta$ (J2000) = $-$08$^{\circ}$ 48' 07.1''
which marks the center of the elliptical component that describes the shape of
the X-ray emission on larger scales.

Figure~\ref{2dxfit} shows the contours of the best-fitting model
overlaid both on the observed exposure-corrected X-ray image and on
the residuals remaining when the model is subtracted from the
data. The residual image shows a clear excess of emission,
corresponding to about 300 photons, to the West of the cluster
core. Note that the best-fit model parameters quoted above were
obtained with the excess region excluded from the fit.

Since the two-dimensional fitting procedure allows no immediate
assessment of the goodness of fit other than via visual inspection of
the residuals (Fig.~\ref{2dxfit}), we also fit a one-dimensional,
spherical $\beta$ model to the radial X-ray surface brightness
profile. In this fit we adopt the center of the elliptical model
component as the overall center of the X-ray emission, and exclude a
60-degree-wide azimuthal section around the compact core and the
excess emission to the West-North-West. Again we account for
variations in the exposure time across the source and background
regions. Since all annuli contain at least 50 photons we are now
justified in using $\chi^2$ statistics in the fit.

\begin{figure}
\epsfxsize=0.5\textwidth
\epsffile{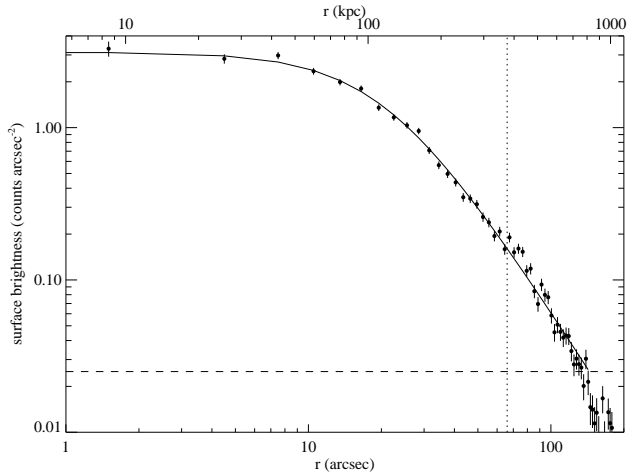}
\caption{\label{1dxfit}Radial X-ray surface brightness profile of
  MACS\,J1206.2$-$0847 in the 0.5--7 keV band as observed with
  Chandra/ACIS-I. The profile is centered on the peak of the
  elliptical surface brightness component as determined in our
  two-dimensional fit (see text for details). The solid line shows the
  best-fitting $\beta$ model. The dashed horizontal line marks the
  surface brightness corresponding to twice the background level; the
  vertical dotted line shows the radial limit within which a $\beta$
  model provides an acceptable description of the data.}
\end{figure}

The resulting one-dimensional radial profile as well as the
best-fitting $\beta$ model are shown in Figure~\ref{1dxfit}. When
fitting out to a radius of about 150 arcsec (850 kpc), where the
observed surface brightness profile begins to drop below twice the
background level, we find the $\beta$ model to provide an unacceptable
fit to the data at a reduced $\chi^2$ value of 1.9 (48 data points, 45
degrees of freedom (d.o.f.). However, essentially the same model fits
the data very well ($\chi^2$=1.1 for 19 d.o.f.) within a radius of 66
arcsec (375 kpc).  At larger radii the observed slope varies such that
the best-fitting $\beta$ model first systematically underpredicts, then
systematically exceeds the observed values. The best-fit values
of $S_0=(3.23\pm 0.13)$ ct arcsec$^2$, $r_0=(20.1\pm 1.3)$ arcsec
(corresponding to ($114\pm 7$) kpc), and $\beta=0.57\pm 0.02$, all of
which are consistent with the results from our two-dimensional fit,
thus allow a credible parametrization of the observed emission out to
about 375 kpc, provided the compact core and the excess emission
region to the West are excluded. At larger radii, the X-ray morphology
of MACS\,J1206.2$-$0847 is again too complex to be described by a
simple $\beta$ model.

Although the one-dimensional model is too simplistic to adapt to the spatial
variations in the X-ray emission near the cluster core or at very large radii,
it provides an adequate global description of the cluster. Extrapolating the
model to $r_{\rm 200}$ (see Section~\ref{xmass}) yields values of $(4.4\pm 0.07)
\times 10^{-12}$ erg s$^{-1}$ cm$^{-2}$ and $(24.3\pm 0.5)\times 10^{44}$ erg
s$^{-1}$ for the total X-ray flux and X-ray luminosity of MACS\,J1206.2--0847,
respectively (0.1--2.4 keV). The quoted errors do, however, not account for any
systematic errors which are bound to be present in view of the fact that X-ray
emission is detected only out to about 1 Mpc from the cluster center and that,
at larger radii, the $\beta$ model tends to overpredict the observed X-ray
surface brightness (see Fig.~\ref{1dxfit}). A more robust measurement can be
obtained within $r_{\rm 1000}$ (1.04 Mpc) and yields lower limits to the total
X-ray flux and luminosity of $(3.8\pm 0.06) \times 10^{-12}$ erg s$^{-1}$
cm$^{-2}$ and $(21.0\pm 0.4)\times 10^{44}$ erg s$^{-1}$.

\subsection{Spectral analysis}
\label{xspec}

\begin{figure}
\epsfxsize=0.5\textwidth
\epsffile{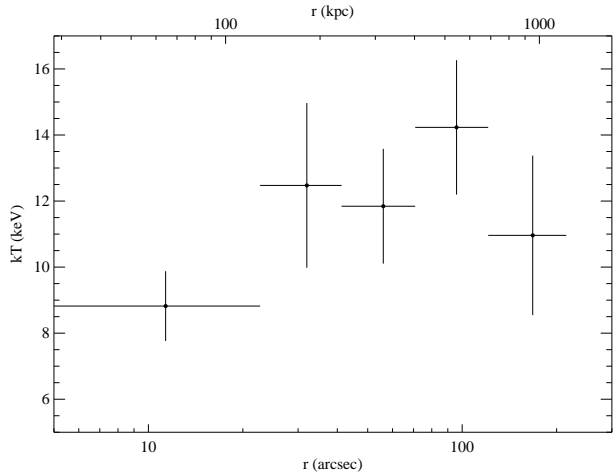}
\caption{\label{tprof}Radial profile of the ICM temperature in
  MACS\,J1206.2$-$0847. The profile is centered on the peak of the
  elliptical surface brightness component as determined in our
  two-dimensional fit (see text for details). Horizontal bars mark the
  width of the respective annulus, defined such that each region
  contains about 3,000 net photons.}
\end{figure}

We measure a global temperature for the intra-cluster medium (ICM) in
MACS\,J1206.2$-$0847 by extracting the X-ray spectrum from $r=70$ kpc
to $r=1$ Mpc (r$_{1000}$) and using \textit{Sherpa} to fit a MEKAL
plasma model (Mewe et al.\ 1985) with the absorption term frozen at
the Galactic value. We find ${\rm k}T= (11.6\pm 0.66)$ keV, a high
value even for extremely X-ray luminous clusters (Chen et
al.\ 2007). Although the relatively high reduced-$\chi^2$ value of
this spectral fit of 1.4 is statistically acceptable, it could be
indicative of systematic effects such as spatial temperature
variations or the presence of multiphase gas. We find only mild
evidence of the former when fitting absorbed plasma models to the
X-ray spectra extracted from five concentric annuli. As shown in
Fig.~\ref{tprof}, the ICM temperature is consistently high ($\sim 12$
keV), with the exception of the core region where, at $r<130$ kpc, a
significant drop to about 9 keV (still a very high value) is
observed. The lower gas temperature measured around the cluster core
could be caused by the presence of a minor cool core, in agreement
with the results of our X-ray imaging analysis.

\section{Properties of the central galaxy}

\begin{figure}
\epsfxsize=0.5\textwidth
\epsffile{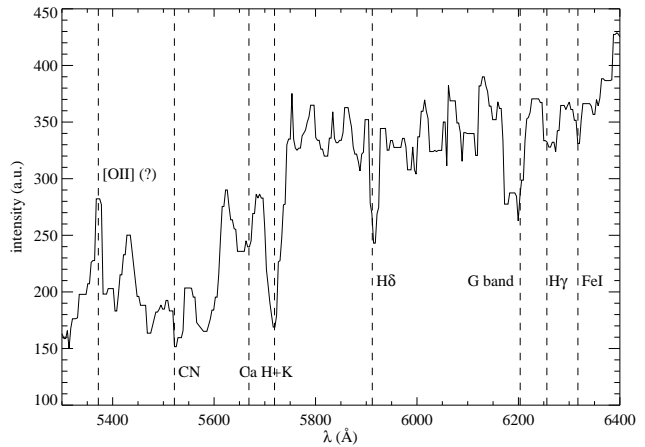}
\caption{\label{cdspec}Low-resolution spectrum of the central cluster
galaxy in MACSJ1206.2$-$0847 as observed with CFHT (see text for
details). A tentative detection of OII in emission is highlighted, as
are a series of absorption features characteristic of the old stellar
population dominating cluster ellipticals.}
\end{figure}

The spectrum of the central cluster galaxy taken at CFHT
(Fig.~\ref{cdspec}, see also Section~\ref{optspec}) covers only a
small wavelength range mostly redward of the 4000\AA\ break.  While
the wavelength coverage is thus insufficient to check for the presence
of H$\beta$ and OIII in emission, we do detect faint OII emission
($\lambda_{\rm rest}=3727$ \AA), albeit at a much lower level than
typically observed in the central cluster galaxies of large cool-core
clusters (e.g., Allen et al.\ 1992, Crawford et al.\ 1995).

\begin{figure}
\epsfxsize=0.5\textwidth
\epsffile{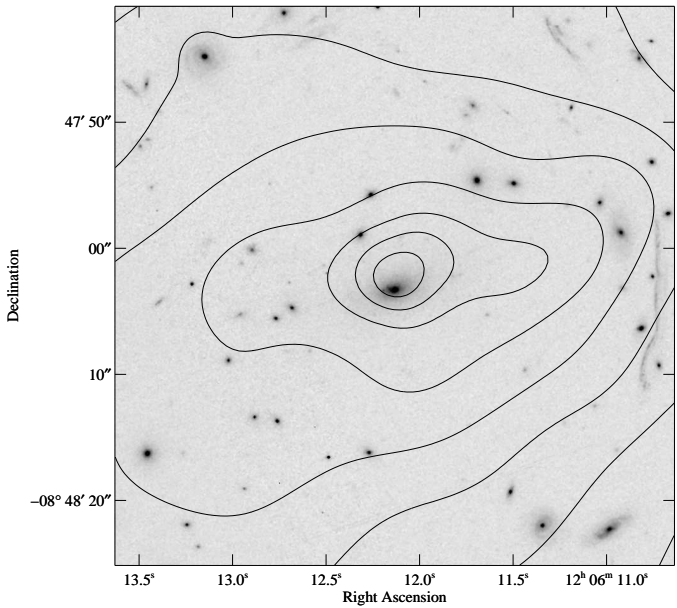}
\caption{\label{corealign}As Figure~\ref{xcont} but using the HST
  image and zoomed in to show only the cluster core.}
\end{figure}

Bright NVSS (162 mJy at 1.4~GHz) and Molonglo (900 mJy at 365~MHz)
radio sources are coincident with the central galaxy. The radio source
exhibits a relatively steep spectrum ($\alpha=-1.32\pm0.05$) and is
similar to the one found in MACSJ1621.3+3810 (Edge et al.\ 2003) but
an order of magnitude more powerful (6$\times 10^{26}$ W~Hz$^{-1}$ at
1.4~GHz). The radio source is also detected at 74 MHz in the VLA
Low-frequency Sky Survey (VLSS) with a measured flux of
$(6.67\pm0.7)$ Jy that is in excellent agreement with the power-law
prediction from the detections at 365 MHz and 1.4 GHz.

Although no clear signs of cavities are detected in the cluster core, we can,
from our Chandra data alone, not conclusively assess the degree of interaction
between the BCG and the surrounding intra-cluster medium. High-resolution radio
data would be required to map the radio morphology of the BCG. We do observe
though a slight, but significant displacement of $(1.7\pm 0.4)$ arcsec --
$(9.6\pm 2.3)$ kpc at the cluster redshift -- between the position of the
central cluster galaxy ($\alpha$ (J2000) = 12$^{\rm h}$ 06$^{\rm m}$ 12.14$^{\rm
  s}$, $\delta$ (J2000) = $-$08$^{\circ}$ 48' 03.3'') and the X-ray centroid of
the compact cluster core (Fig.~\ref{corealign}). Having carefully aligned the
optical and X-ray images (see footnote in Section~\ref{gas}) we estimate that at
most 10\% of this offset can reasonably be attributed to residual astrometric
uncertainties; the majority of the misaligned is thus real. Similar offsets
($\sim$ 10 kpc) have been noted in previous Chandra studies of galaxy clusters
(Arabadjis, Bautz \& Garmire 2002) and may also have contributed to minor
optical/X-ray misalignments in the cores of cooling clusters observed with ROSAT
(Peres et al.\ 1998).

In addition, the velocity offset of 550 km s$^{-1}$ between the
central galaxy and the cluster mean velocity is comparable to the
largest peculiar velocities observed in local clusters (Zabludoff et
al.\ 1990; Hill \& Oegerle 1993; Oegerle \& Hill 2001), although we
note that the non-Gaussian velocity distribution in this cluster
(Fig.~\ref{zhist}) complicates the measurement of an accurate systemic
velocity.

Spatial and velocity offsets between central galaxy and X-ray peak are
not expected in the simplest cool-core scenario, although it is worth
noting that an offset of the same size as observed by us here (10 kpc)
has been seen in the best-studied cool-core cluster, Perseus
(B\"ohringer et al.\ 1993), and there has been a widespread
realization that the physics of cool cluster cores are more complex
and the role of AGN feedback more important than previously thought
(Peterson et al.\ 2001; Soker et al.\ 2002; Edge 2001; Mittal et
al.\ 2008). We conclude that MACSJ1206.2$-$0847 is likely to contain a
moderate cool core, as well as an extremely luminous radio galaxy (one
of the most powerful ones known in cluster cores at $z>0.4$), with the
observed disturbances being likely due to a recent or still ongoing
cluster merger.

\section{Mass measurements}

\subsection{Virial mass}

Using the measured redshifts of cluster members and their spatial distribution
as projected on the sky, we determined the virial cluster mass based on the
method of Limber \& Mathews (1960) in which the mass is calculated as
\begin{equation}
M_V = \frac{3\pi}{2} \frac{\sigma_{P}^{2}R_H}{G}.
\label{M_Veqn}
\end{equation}

Here $\sigma_P$ is the one-dimensional (radial) velocity dispersion and $R_H$ is
the projected mean harmonic point-wise separation (projected virial
radius). $R_H$ is defined by
\begin{equation}
R_H^{-1} = \frac{1}{N^2}\sum_{i<j} \frac{1}{|\mathbf{r}_i-\mathbf{r}_j|},
\label{R_Heqn}
\end{equation}

where N is the number of galaxies, $|\mathbf{r}_i-\mathbf{r}_j|$ is the
projected separation of galaxies $i$ and $j$, and the $ij$ sum is over all
pairs.  Being a pairwise estimator this quantity is sensitive to close pairs and
quite noisy (Bahcall \& Tremaine 1981).  It also systematically underestimates
the radius for a rectangular aperture typical of cluster redshift surveys
(Carlberg et al.\ 1996).  Carlberg et al.\ therefore introduce a new radius
estimator, the ringwise projected harmonic mean radius $R_h$, given by
\begin{eqnarray}
R_h^{-1} &=& \frac{1}{N^2} \sum_{i<j}\frac{1}{2\pi} \int_{0}^{2\pi}\frac{d\theta}{\sqrt{r_{i}^{2}+r_{j}^{2}+2r_ir_j\cos\theta}} \nonumber \\ 
&=& \frac{1}{N^2} \sum_{i<j}\frac{2}{\pi (r_i+r_j) K(k_{ij})}.
\label{R_heqn}
\end{eqnarray}

Here $r_i$ and $r_j$ are the projected distances from the cluster center to
galaxies $i$ and $j$ respectively, $k_{ij}^{2} = 4r_ir_j/(r_i+r_j)^2$, and
$K(k)$ is the complete elliptic integral of the first kind in Legendre's
notation.  This estimator requires an explicit choice of cluster center and
assumes the cluster is spherically symmetric with respect to this center.  It
treats one of the particles in the pairwise potential $|r_i-r_j|^{-1}$ as having
its mass distributed in a ring around the cluster center. $R_h$ is less
sensitive to close pairs, less noisy, and tolerates non-circular apertures
better than $R_H$.  If the cluster is significantly flattened or subclustered,
however, $R_h$ will systematically overestimate the true projected virial
radius.

We calculate both radius estimators in our mass determinations to investigate
the resulting systematic uncertainty in the virial mass. For our sample the
virial radius and mass derived using $R_h$ are 7\% larger than those based on
$R_H$. We choose to use $R_h$ as the more robust estimator in our analysis and
define the three-dimensional (deprojected) virial radius as
\begin{equation}
r_V = \frac{\pi}{2}R_h.
\label{r_Veqn}
\end{equation}

Our determination of the virial radius estimators $R_H$ and $R_h$ was made using
all 62 galaxies with redshifts within $3\sigma$ of the cluster mean (see
Section~\ref{galprop}).  The resulting projected virial radius is $R_h = 1.176$
Mpc ($R_H = 1.096$ Mpc) with a virial mass of $3.861\times10^{15}$ M$_{\odot}$
and a three-dimensional virial radius of $r_V = 1.847$ Mpc.

\subsection{X-ray mass}
\label{xmass}
To estimate the total gravitational mass of the cluster from its X-ray emission,
we need to assume that the cluster is in hydrostatic equilibrium. In addition,
we need a description of the density as well as the temperature of the
intracluster gas, often assumed to be isothermal. For a cluster with a core
region as disturbed as the one of MACSJ1206.2$-$0847 such simplifying
assumptions are unlikely to be justified; however, an isothermal $\beta$ model
should provide an adequate description of the cluster outskirts and allow us to
obtain a crude mass estimate.

From the X-ray temperature (see Section~\ref{xspec}) we estimate the virial
radius $R_{200}$ using the formula of Arnaud et al.\ (2002),
\begin{eqnarray*}
R_{200} & = & 3.80~\beta_T^{1/2}\Delta_z^{-1/2}~(1+z)^{-3/2} \nonumber \\
        &  & \times(\frac{kT}{10{\rm keV}})^{1/2}h_{50}^{-1}~{\rm Mpc} \nonumber
\end{eqnarray*} 
with
\begin{eqnarray*}
\Delta_z  =  (200\Omega_0)/(18\pi^2\Omega_z). \nonumber
\end{eqnarray*}
Here $\beta_T=1.05$ (Evrard et al.\ 1996) is the normalization of the virial relation, i.e.,
$GM_v/(2R_{200})= \beta_T~{\rm k}T$.  Then, the total mass of the
cluster within a radius $r$ can be computed with the help of the $\beta$-model profile
discussed in Section~\ref{xbeta}:
\begin{eqnarray}
M(r) & = & 1.13\times10^{14}\beta\frac{T}{\rm{keV}}\frac{r}{\rm{Mpc}}\frac{(r/r_c)^2}{1+
(r/r_c)^2}M_{\odot}
\end{eqnarray}
(Evrard et al.\ 1996). Using the above equations we find $R_{\rm 200}=(2.3\pm
0.1)$ Mpc as an approximate value for the virial radius. The X-ray estimates for
the total mass within $R_{\rm 200}$ and the mass within the core region (defined
as the sphere interior to the giant arc, i.e.\ $r<119$ kpc) are then $(17.1\pm
1.2)\times 10^{14}$ M$_{\sun}$ and $(0.46\pm 0.05)\times 10^{14}$~M$_{\sun}$,
respectively.

\subsection{Lensing mass}

\begin{figure}
\epsfxsize=0.5\textwidth
\epsffile{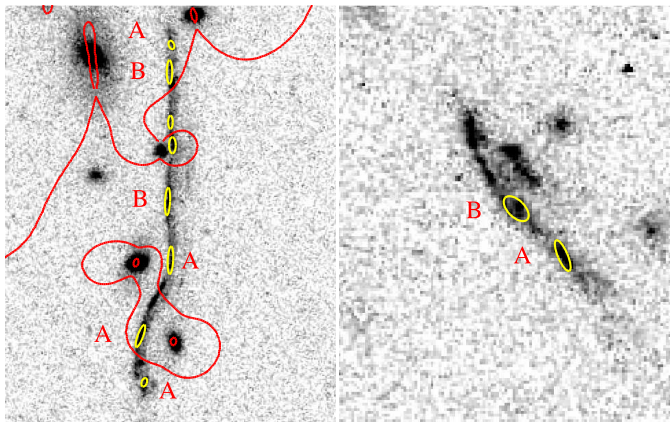}
\caption{ \label{fig:critics_closeup} Close-up of the critical lines
  at the positions of the giant arc and the counter image. The yellow
  ellipses and identifiers mark components of the set of multiple
  images used to constrain the lens model. }
\end{figure}

Using the high-resolution HST images, we have identified within the
giant arc two features A \& B that are replicated six times. The same
two features are also identified in the southern part of the counter
image as shown in Figure~\ref{fig:critics_closeup}. The northern part
of the counter image is most likely not multiply imaged. Altogether,
the giant arc and its counter image represent a seven-image multiple
system which we use to constrain a strong-lensing model of the cluster
mass distribution.

\begin{figure}
\epsfxsize=0.5\textwidth
\epsffile{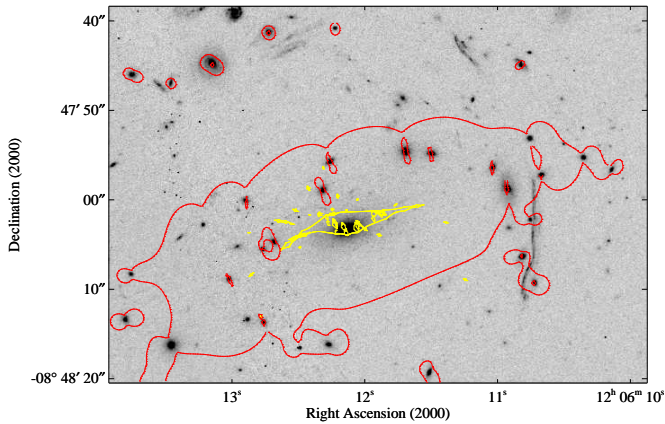}
\caption{\label{fig:critics_core} HST/ACS image of the cluster core
  observed through the F606 filter. The critical line computed at the
  redshift of the giant arc ($z=1.036$) is shown in red. The yellow
  curve shows the corresponding caustics in the source plane. }
\end{figure}

To model the cluster core we used \textsc{Lenstool}\footnote{publicly
  available at http://www.oamp.fr/cosmology/lenstool} (Kneib et
al.\ 1996; Jullo et al.\ 2007) which now uses a Bayesian MCMC sampler
to optimize the cluster mass model and generate robust lens results.
We have followed the procedure of Limousin et al.\ (2007) to model the
mass distribution using one cluster-scale dark-matter halo described
by a truncated PIEMD (Pseudo Isothermal Elliptical Mass Distribution),
as well as an additional 84 truncated galaxy-scale PIEMD potentials to
describe the dark-matter halos associated with the brightest cluster
member galaxies selected from the cluster V--K red
sequence. Furthermore, to minimize the number of free parameters, we
assumed that the mass of galaxy-size halos scales with the K-band
luminosity of the associated galaxy (Natarajan \& Kneib 1997). The
obtained critical curves shown in figure~\ref{fig:critics_core}
display a winding shape in between the galaxies (see close-up in
Fig.~\ref{fig:critics_closeup}), which explains the extreme elongation
of the giant arc.

Using the most probably strong-lensing mass model, we estimate the mass
enclosed by the giant arc. We find $M(<21'') = (112.0 \pm 5) \times
10^{12}\ M_{\odot}$ and a mass-to-light ratio interior to the giant
arc of $M/L = 56\pm2.5$. 

These numbers are very robust and depend little on the mass profile
assumed for the dark-matter distribution of the cluster. Caution is
advised though when extrapolating to larger radius, as the slope of
the cluster mass profile is not well constrained by only one
(multiple) arc. Deeper, high-resolution imaging (with, e.g., ACS or
WFC3), however, would likely detect a large number of multiple images
as was the case for Abell~1703 (Limousin et al.\ 2008), allowing us to
accurately measure the slope of the cluster dark-matter profile. Note
that the current best estimate of the Einstein radius at $z\sim 7$ is
nearly 45\arcsec, making this cluster a superb cosmological telescope
to probe the first galaxies in the Universe.

The total magnification of the system (both the giant arc and the part
of the counter image that is multiply imaged) is about 80$\pm$10, one
of the largest amplification factors known for a giant arc. A detailed
model of the arc surface brightness is beyond the scope of this study
but will be presented in a future paper (Clement et al., in
preparation).

\section{Summary and Conclusions}

We present a comprehensive multi-wavelength analysis of the properties of the
massive galaxy cluster MACS\,J1206.2--0847. At a redshift of $z=0.4385$, the
system acts as a gravitational lens for a background galaxy at $z=1.04$,
resulting in spectacular gravitational arc of high surface brightness, 15\arcsec
in length, a total magnitude of V=21.0, and of unusual, very red colour of
$R-K=4.3$. Our X-ray analysis based on Chandra data yields global X-ray
properties ($L_{\rm X}=2.3\times 10^{45}$ erg s$^{-1}$, 0.1--2.4 keV, and ${\rm
  k}T=11.6\pm 0.7$ keV) that make this cluster one of the most extreme systems
known at any redshift.

Belying its relaxed appearance at optical wavelengths, MACS\,J1206.2--0847
exhibits many signs of ongoing merger activity along the line of sight when
looked at more closely, including a disturbed X-ray morphology in the cluster
core, a small but significant offset of the peak of the X-ray emission from the
brightest cluster galaxy, and a very high velocity dispersion of 1580 km
s$^{-1}$. The strongest indication of recent or ongoing cluster growth, however,
is obtained from a comparison of X-ray, virial, and lensing mass estimates for
this system. A high-resolution image of the giant arc and its counter image
obtained with HST allows us to create a lens model that places tight constraints
on the mass distribution interior to the arc. The strong-lensing value of the
mass of the cluster core of $(11.2 \pm 0.5) \times 10^{13}\ M_{\odot}$ thus
obtained is higher by about a factor of two than the X-ray estimate of $(4\pm
0.4)\times10^{13}$ M$_{\odot}$. A similar discrepancy is found between the X-ray
estimate of the total mass within $r_{\rm 200}$ and the virial mass estimate
derived from radial-velocity measurements for 38 cluster galaxies.

Comparable discrepancies between X-ray and lensing mass estimates, in particular
for cluster cores, have been reported before for other systems, the perhaps most
famous example being A1689 (e.g., Miralda-Escud\'e \& Babul 1995; Xue \& Wu
2002; Limousin et al.\ 2007). In all cases, including the one presented here,
the mass derived using gravitational-lensing features is two to three times
higher than the one obtained by an X-ray analysis assuming hydrostatic
equilibrium. In agreement with simulations (Bartelmann \& Steinmetz 1996),
detailed observational studies of such discrepancies for individual clusters
find deviations from hydrostatic equilibrium and the presence of substructure
along the line of sight to be responsible, tell-tale signs being offsets between
the X-ray peak and the location of the BCG as well as extreme elongations and
structure in radial-velocity space (Allen 1998, Machacek et al.\ 2001). We
conclude that MACS\,J1206.2--0847 is a merging cluster with a merger axis that
is close to aligned with our line of sight. A modest cool core has either
survived the merger or is in the process of formation.

The discovery of a giant arc in this MACS cluster underlines yet again the
efficiency of X-ray luminous clusters as gravitational lenses. Much deeper,
high-resolution images of systems like MACS\,J1206.2--0847 will, owing to the
large number of multiple images detected, allow a detailed mapping of the mass
distribution in the cluster core, and, through the power of gravitational
magnification, provide an ultra-deep look at the very distant Universe.

\section*{Acknowledgments}

We thank Alexey Vikhlinin for very helpful advice and suggestions
concerning systematic effects in the Chandra data analysis.  HE
gratefully acknowledges financial support from grants NAG 5-8253 and
GO2-3168X. ACE thanks the Royal Society for generous support during
the gestation phase of this paper.  JPK acknowledges support from the
{\it Centre National de la Recherche Scientifique} (CNRS), and the ANR
grant 06-BLAN-0067.  EJ acknowledges support from an ESO-studentship
and from CNRS.

\begin{table*}
\caption{Positions, cross-correlation peak heights, and spectroscopic
  redshifts (heliocentric) of galaxies in the field of
  MACS\,J1206.2$-$0847. The origin of each measurement (FORS1 on the
  VLT or MOS on CFHT) is marked in the final column. \label{ztbl}}
\begin{tabular}{lccccc}
\hline
R.A. (J2000) & Dec (J2000) & Peak Height & $z$ & $\Delta z$  & Telescope\\
\hline
12 05 58.37  &  $-$08 47 48.5  &  0.88   &  0.44255  &  0.00003  &  CFHT \\
12 05 59.28  &  $-$08 49 34.8  &  0.47   &  0.44921  &  0.00024  &  CFHT \\
12 06 01.42  &  $-$08 49 31.7  &  0.46   &  0.44378  &  0.00031  &  CFHT \\
12 06 02.48  &  $-$08 44 46.4  &  0.76   &  0.44669  &  0.00013  &  CFHT \\
12 06 03.29  &  $-$08 46 18.7  &  0.58   &  0.44081  &  0.00144  &  CFHT \\
12 06 03.31  &  $-$08 48 38.0  &  0.92   &  0.43766  &  0.00005  &  CFHT \\
12 06 04.25  &  $-$08 51 09.4  &  0.49   &  0.42341  &  0.00108  &  CFHT \\
12 06 04.72  &  $-$08 46 12.8  &  0.79   &  0.34227  &  0.00012  &  CFHT \\
12 06 05.27  &  $-$08 46 27.5  &  0.83   &  0.52920  &  0.00005  &  CFHT \\
12 06 05.45  &  $-$08 49 05.5  &  0.72   &  0.43627  &  0.00006  &  CFHT \\
12 06 05.71  &  $-$08 50 31.7  &  0.81   &  0.43547  &  0.00005  &  CFHT \\
12 06 05.91  &  $-$08 48 13.1  &  0.88   &  0.43505  &  0.00009  &  CFHT \\
12 06 06.27  &  $-$08 50 41.2  &  0.79   &  0.43616  &  0.00010  &  CFHT \\
12 06 06.47  &  $-$08 48 48.7  &  0.88   &  0.44432  &  0.00006  &  VLT  \\
12 06 06.75  &  $-$08 44 27.5  &  0.79   &  0.43563  &  0.00009  &  CFHT \\
12 06 06.85  &  $-$08 50 46.4  &  0.81   &  0.42495  &  0.00004  &  CFHT \\
12 06 06.99  &  $-$08 49 08.4  &  0.88   &  0.43461  &  0.00003  &  VLT  \\
12 06 07.23  &  $-$08 48 45.1  &  0.77   &  0.45409  &  0.00006  &  CFHT \\
12 06 07.76  &  $-$08 47 18.0  &  0.95   &  0.43245  &  0.00003  &  VLT  \\
12 06 07.85  &  $-$08 50 17.7  &  0.82   &  0.59625  &  0.00006  &  CFHT \\
12 06 08.15  &  $-$08 48 24.6  &  0.79   &  0.43371  &  0.00007  &  CFHT \\
12 06 08.60  &  $-$08 50 09.5  &  0.64   &  0.44649  &  0.00012  &  CFHT \\
12 06 08.62  &  $-$08 45 04.0  &  0.71   &  0.44152  &  0.00010  &  VLT  \\
12 06 08.78  &  $-$08 47 00.0  &  0.90   &  0.44008  &  0.00002  &  VLT  \\
12 06 09.54  &  $-$08 47 31.1  &  0.88   &  0.45178  &  0.00005  &  CFHT \\
12 06 09.66  &  $-$08 50 44.4  &  0.34   &  0.36688  &  0.00021  &  VLT  \\
12 06 09.76  &  $-$08 47 19.2  &  0.53   &  0.43981  &  0.00009  &  CFHT \\
12 06 10.71  &  $-$08 47 57.2  &  0.89   &  0.44486  &  0.00004  &  VLT  \\
12 06 10.73  &  $-$08 45 30.0  &  0.79   &  0.68794  &  0.00005  &  VLT  \\
12 06 10.76  &  $-$08 48 09.4  &  0.75   &  0.44460  &  0.00008  &  VLT  \\
12 06 10.96  &  $-$08 49 26.9  &  0.82   &  0.43782  &  0.00007  &  VLT  \\
12 06 11.00  &  $-$08 46 27.7  &  0.84   &  0.43840  &  0.00004  &  VLT  \\
12 06 11.02  &  $-$08 50 22.9  &  0.96   &  0.42781  &  0.00003  &  VLT  \\
12 06 11.76  &  $-$08 47 55.5  &  0.71   &  0.42274  &  0.00012  &  CFHT \\
12 06 12.20  &  $-$08 48 04.1  &  0.71   &  0.44132  &  0.00006  &  CFHT \\
12 06 13.00  &  $-$08 51 04.0  &  0.55   &  0.43312  &  0.00024  &  CFHT \\
12 06 13.04  &  $-$08 47 40.1  &  0.54   &  0.53602  &  0.00017  &  CFHT \\
12 06 13.18  &  $-$08 47 44.8  &  0.78   &  0.42373  &  0.00006  &  VLT  \\
12 06 13.34  &  $-$08 47 37.4  &  0.70   &  0.42762  &  0.00012  &  CFHT \\
12 06 13.71  &  $-$08 51 01.3  &  0.38   &  0.35605  &  0.00021  &  VLT  \\
12 06 14.04  &  $-$08 47 11.8  &  0.52   &  0.43725  &  0.00010  &  CFHT \\
12 06 14.90  &  $-$08 48 16.3  &  0.70   &  0.42653  &  0.00009  &  CFHT \\
12 06 15.02  &  $-$08 50 34.0  &  0.85   &  0.43421  &  0.00007  &  CFHT \\
12 06 15.71  &  $-$08 48 22.4  &  0.61   &  0.44184  &  0.00013  &  CFHT \\
12 06 16.14  &  $-$08 48 06.9  &  0.85   &  0.44138  &  0.00004  &  CFHT \\
12 06 17.28  &  $-$08 48 23.7  &  0.93   &  0.44498  &  0.00006  &  CFHT \\
12 06 18.06  &  $-$08 49 03.6  &  0.62   &  0.53216  &  0.00011  &  CFHT \\
12 06 18.36  &  $-$08 47 26.8  &  0.37   &  0.44816  &  0.00084  &  CFHT \\
12 06 19.23  &  $-$08 47 46.2  &  0.86   &  0.43826  &  0.00005  &  CFHT \\
12 06 19.59  &  $-$08 45 35.2  &  0.92   &  0.44702  &  0.00006  &  CFHT \\
12 06 19.70  &  $-$08 47 47.9  &  0.86   &  0.44760  &  0.00003  &  CFHT \\
12 06 20.53  &  $-$08 45 34.9  &  0.34   &  0.43464  &  0.00077  &  CFHT \\
12 06 20.73  &  $-$08 50 40.4  &  0.84   &  0.53072  &  0.00005  &  CFHT \\
12 06 21.97  &  $-$08 48 31.5  &  0.43   &  0.43494  &  0.00017  &  CFHT \\
12 06 22.36  &  $-$08 48 42.2  &  0.77   &  0.44310  &  0.00018  &  CFHT \\
12 06 22.84  &  $-$08 46 04.1  &  0.81   &  0.43793  &  0.00013  &  CFHT \\
12 06 23.36  &  $-$08 49 56.1  &  0.64   &  0.43830  &  0.00008  &  CFHT \\
12 06 23.78  &  $-$08 45 17.6  &  0.77   &  0.59400  &  0.00007  &  CFHT \\
12 06 24.40  &  $-$08 50 42.6  &  0.82   &  0.44680  &  0.00011  &  CFHT \\
12 06 24.70  &  $-$08 48 41.3  &  0.80   &  0.43520  &  0.00019  &  CFHT \\
12 06 25.74  &  $-$08 45 34.5  &  0.73   &  0.60783  &  0.00018  &  CFHT \\
12 06 26.04  &  $-$08 50 14.7  &  0.83   &  0.44389  &  0.00002  &  CFHT \\
\hline
\end{tabular}
\end{table*}

\clearpage


\begin{thebibliography}{99}
\bibitem{b0} Allen S.W., 1998, MNRAS, 296, 392
\bibitem{b1} Allen S.W.\ et al. 1992, MNRAS, 259, 67
\bibitem{b2} Allen S.W., Schmidt R.W., Fabian A.C., Ebeling H. 2003, MNRAS,
	342, 287
\bibitem{b3} Arabadjis J.S., Bautz M.W., Garmire G.P. 2002, ApJ, 572, 66
\bibitem{b4} Arag\'{o}n-Salamanca A.\ \& Ellis R.S. 1990, in 
	Mellier Y., Soucail G., Fort B., ed, Gravitational Lensing. 
	Springer Verlag, p.\ 288
\bibitem{b5} Arnaud M., Aghnaim N., \& Neumann D. 2002, A\&A, 389, 1
\bibitem{b6} Bahcall J.\ \& Tremaine S.D. 1981, ApJ, 447, L81
\bibitem{b6a} Bartelmann M.\ \& Steinmetz M. 1996, MNRAS, 283, 431
\bibitem{b7} Beers T.C., Flynn K., Gebhardt K. 1990, AJ, 100, 32
\bibitem{b8} B\"ohringer H., Voges W, Fabian A.C., Edge A.C., Neumann D.M.
	1993, MNRAS, 264, L25
\bibitem{b9} Borgani S.\ et al. 2001, ApJ, 561, 13
\bibitem{b10} Cavaliere A.\ \& Fusco-Femiano R. 1976, A\&A, 49, 137
\bibitem{b11} Chen Y., Reiprich T.H., B\"ohringer H., Ikebe Y., Zhang Y.-Y., 2007, A\&A 466, 805
\bibitem{b12} Carlberg R.G., Yee H.K.C., Ellingson E., Abraham R., 
        Gravel P., Morris S.\ \& Pritchet C.J. 1996, ApJ, 462, 32
\bibitem{b13} Cash W. 1979, ApJ, 228, 939
\bibitem{b14} Covone G., Kneib J.-P., Soucail G., Richard J., Jullo E., Ebeling H. 2006, A\&A, 456, 409
\bibitem{b15} Crawford C.S., Edge A.C., Fabian A.C., Allen S.W., B\"ohringer H.,
        Ebeling H., McMahon R.G., Voges W. 1995, MNRAS, 274, 75
\bibitem{b16} Dahle H., Kaiser N., Irgens R.J., Lilje P.B., Ridgway S.E.  2002,
        ApJS, 139, 313
\bibitem{b17} Ebeling H., Edge A.C., Henry J.P. 2001, ApJ, 553, 668
\bibitem{b18} Ebeling H., White D.A., Rangarajan F.V.N. 2006, MNRAS, 368, 65
\bibitem{b19} Ebeling H., Barrett E., Donovan D., Ma C.-J., Edge A.C., van Speybroeck, L. 2007 ApJ, 661, L33
\bibitem{b20} Edge A.C. 2001, MNRAS, 328, 762
\bibitem{b21} Edge A.C., Ebeling H., Bremer M., R\"ottgering H., van Haarlem M.,
	Rengelink R., Courtney N. 2003, MNRAS, 339, 913
\bibitem{b22} Evrard A.E., Metzler C.A., \& Navarro J.F. 1996, ApJ, 469, 494
\bibitem{b23} Gioia I.M., Luppino G.  1994, ApJS, 94, 583
\bibitem{b24} Henry J.P. 2000, ApJ, 534, 565
\bibitem{b25} Hill J.M.\ \& Oegerle W.R. 1993, AJ, 106, 831
\bibitem{b26} Jullo E., Kneib J.-P., Limousin M., El\'{i}asd\'{o}ttir \'{A}., Marshall P.J., Verdugo T. 2007, NJPh, 9, 447
\bibitem{b28} Kneib J.-P., Ellis R.S., Smail I., Couch W.J., Sharples R.M. 1996, ApJ, 471, 643
\bibitem{b29} Kneib J.-P., Ellis R.S., Santos M.R., Richard J. 2004, ApJ, 607, 697
\bibitem{b30} Limber D.N.\ \& Mathews W.G. 1960, ApJ, 132, 286
\bibitem{b31} Limousin M., Kneib J.-P., Bardeau S., Natarajan P., Czoske O., Smail I., Ebeling H., Smith G.P. 2007, A\&A, 461, 881
\bibitem{b31b} Machacek M., Bautz M.W., Canizares C., Garmire G.P. 2002, ApJ, 567, 188
\bibitem{b32} Markevitch M.\ \& Vikhlinin A. 2001, ApJ, 563, 95
\bibitem{b33} Mellier Y., Fort B., Soucail G., Mathez G., Cailloux, M. 1991, ApJ, 380, 334
\bibitem{b34} Mewe R., Gronenschild E.H.B.M., \& van den Oord G.H.J. 1985, A\&AS, 62, 197
\bibitem{b34a} Mittal R. Hudson D.S., Reiprich T.H., Clarke T. 2008, A\&A, submitted, arXiv:0810.0797
\bibitem{b34b} Miralda-Escud\'e J.\ \& Babul A. 1995, ApJ, 449, 18
\bibitem{b35} Natarajan P.\ \& Kneib J.-P. 1997, MNRAS, 287, 833
\bibitem{b36} Navarro J.F., Frenk C.S., White S.D.M 1997, ApJ, 490, 493
\bibitem{b37} Oegerle W.R.\ \& Hill J.M. 2001, AJ, 122, 2858
\bibitem{b38} Peterson J.R.\ et al. 2001, A\&A, 365, 104L
\bibitem{b39} Pierpaoli E., Borgani S., Scott D., White M. 2003, MNRAS, 342, 163
\bibitem{b40} Peres C.B., Fabian A.C., Edge A.C., Allen S.W., Johnstone R.M.,
	White D.A. 1998, MNRAS, 298, 416
\bibitem{b41} Sand D.J., Treu T., Smith G.P., Ellis R.S. 2003, ApJ, submitted
\bibitem{b42} Smail I., Ellis R.S., Arag\'{o}n-Salamanca A., Soucail G.,
	Mellier Y., Giraud E. 1993, MNRAS, 263, 628
\bibitem{b42} Smith G.P., Kneib J.-P., Ebeling H., Czoske O., Smail I. 2003, 
        ApJ, 552, 493
\bibitem{b44} Smith G.P., Smail I., Kneib J.-P., Davis, C.J., Takamiya, M.
        Ebeling H., Czoske O. 2002, MNRAS, 333, L13
\bibitem{b45} Soker N., Blanton E.L., Sarazin C.L. 2002, ApJ, 573, 533
\bibitem{b46} Soucail G., Mellier Y., Fort B., Mathez G., Hammer
        F. 1987, A\&A, 184, L7
\bibitem{b47} Vikhlinin A., van Speybroeck L., Markevitch M., Forman W.R., 
	Grego L. 2002, ApJ, 578, L107
\bibitem{b47a} Xue S.-J.\ \& Wu X.-P. 2002, ApJ, 576 152
\bibitem{b48} Zabludoff A.I., Huchra J.P., Geller M.J. 1990, ApJS, 74, 1
\end{thebibliography}
\end{document}